\documentstyle[12pt]{article}
\newcommand{\beq}{\begin{equation}}
\newcommand{\eeq}{\end{equation}}

\newcommand{\G}{\Gamma}
\newcommand{\rh}{\rho}
\newcommand{\sg}{\sigma}
\newcommand{\Sg}{\Sigma}
\newcommand{\pr}{\partial}

\renewcommand{\l}{\lambda}
\renewcommand{\L}{\Lambda}
\renewcommand{\b}{\beta}
\renewcommand{\a}{\alpha}
\newcommand{\n}{\nu}
\newcommand{\m}{\mu}

\newcommand{\eps}{\epsilon}

\begin{document}
\topmargin 0pt
\oddsidemargin 5mm
\headheight 0pt
\headsep 0pt
\topskip 5mm

\begin{flushright}
CCNY-HEP-94-3\\
\hfill
April 1994
\end{flushright}

\begin{center}
\hspace{10cm}

\vspace{48pt}
{\large \bf
EXTRINSIC CURVATURE DEPENDENCE OF NIELSEN-OLESEN STRINGS }
\end{center}

\vspace{30pt}

\begin{center}
{\bf Peter Orland}

\vspace{20pt}
The City College of the City University of New York, \\
New York, NY 10031

\vspace{10pt}
and
\vspace{10pt}

Baruch College, \\
The City University of New York, \\
New York, NY 10010\\
orlbb@cunyvm.cuny.edu \footnote{Permanant Address}

\end{center}

\vspace{40pt}

\begin{center}
{\bf Abstract}

\end{center}
It is shown how to
treat the degrees of freedom of Nielsen-Olesen
vortices in the $3+1$-dimensional $U(1)$ higgs model by
a collective coordinate method. In the london limit, where
the higgs mass becomes infinite, the gauge
and goldstone degrees of freedom are integrated out, resulting
in the vortex world-sheet action. Introducing
an ultraviolet cut-off mimics the effect
of finite higgs mass. This action is non-polynomial
in derivatives and
depends on the extrinsic curvature
of the surface. Flat surfaces are stable if the
coherence length is less than the penetration depth. It
is argued that in the quantum abelian higgs model, vortex
world-sheets are dominated by branched
polymers.

\vfill
\newpage

\section{Introduction}

The first topological soliton found in relativistic
gauge theory was Nielsen
and Olesen's string
vortex solution \cite{no}. This solution is essentially
similar to the non-relativistic vortex found by
Abrikosov \cite{abrikosov}. Further work on the structure
of non-abelian vortex
strings was done subsequently by Ezawa and Tze \cite{et}.

Shortly after vortex string solitons were discovered in relativistic
gauge theories, F\"{o}rster \cite{forster} and Gervais and
Sakita \cite{bunji} studied their quantum dynamics. Their analyses
were performed in the limit that both the penetration depth and
coherence length were very small, and concluded that, in that limit, the
vortex is described quite well by a Nambu action.

Since Polyakov \cite{polyakov} discussed the addition of terms in the
string action depending on the extrinsic geometry of the string
world-sheet, a number of people have attempted to determine
whether such terms exist for the Nielsen-Olesen
vortex \cite{A}, \cite{A'}, \cite{B}. In some early references
\cite{A}, \cite{A'}, there was a
consensus that there was a rigidity term in the action
of the vortex, but the sign
of the leading term differed
in these two papers. The most recent word in the literature
\cite{B} was that no extrinsic curvature dependence
seems to be present after all (or if it is present, it is numerically
small). The
basic method used in references \cite{A}, \cite{A'}, \cite{B} is a version
of classical perturbation theory.

The presence of terms quadratic in the
extrinsic curvature tensor in the string
action \cite{polyakov}, \cite{peliti} do not
appear to stabilize the ``branched polymer" disease of quantum
strings \cite{adfo}. This
is because such terms are irrelevant
in the infrared. Polyakov suggested that this may not
be so if another
term, which counts the number of points in which
the world-sheet intersects itself, modulo two, is
included in the action (this term
had been written down earlier by Balachandran et. al. \cite{bal}). It
appears that even classically, certain solutions
are unstable \cite{bz}. Nonetheless, the
presence of such terms may have
implications for cosmic strings \cite{kibble}, which
are essentially classical
Nieslen-Olesen vortices (these strings are presumably
coherent states with energies far above the ground state). The mechanism
for the formation of topological defects
proposed by Kibble \cite{kibble}
has recently been verified in the context of
liquid crystal systems \cite{bowick}.

In this paper, the world-sheet action of the Nielsen-Olesen
vortex is calculated
in the london limit, where the higgs mass
becomes infinite. The calculation
is equally valid for the classical
or quantum case. The method
is closely
related to a representation
of the lattice higgs model \cite{polikarpov}. It is
argued that the effect
of a finite higgs mass is taken into account by
the introduction of an ultraviolet cut-off. This
cut-off is then the reciprocal of the coherence length. The result
is an action analogous to that of
Rasetti and Regge, Davis and Shellard
and of Lund \cite{lr} for liquid helium. However, this form
of the world-sheet action does not make the
extrinsic geometric properties completely clear. The result obtained
disagrees with all of the previous references \cite{A}, \cite{A'}
and \cite{B}. To
leading order in the extrinsic
curvature, the string is found to be anti-rigid (though
evidence is found that the action is stable if
higher orders are taken into account). The result
found in reference \cite{A} was similar, but these
authors found another term at leading order, which is not
revealed by the analysis here.

Effective action methods for vortex dynamics has been criticized
by Arodz and Wegrzyn \cite{arodz}. They pointed out that integrating
out some degrees of freedom leads to a non-local action, which may
not correctly capture the features
of the original model. The essential difficulty is due to
the fact that green's functions depend on the initial
and final conditions. This problem is circumvented
in this paper by going to euclidean space, and considering
finite string manifolds. The basic results (stability
for penetration depth greater than coherence length, no
scaling of the string tension
in the quantum theory) can then be carried
back to Minkowski space.

The action of the vortex in the london limit is non-polynomial in
derivatives on the world-sheet (thus non-local). While a proof
is still lacking, this action appears to be
stable. Small fluctuations about flat surfaces are
examined. The action of a long thin tube is calculated, and it
is argued that in the quantum
theory branched polymers will dominate the world-sheet. This means
that the vortex
string tension probably does not scale for any choice
of couplings of lattice electrodynamics coupled to a higgs
field.

\section{Collective Coordinates}

The starting point is the abelian higgs model in euclidean
space, with lagrangian
\beq
{\cal L}=\frac{1}{4e^{2}}F_{\m \n} F^{\m \n}+\frac{1}{2}
({\pr}_{\n}+iA_{\n})\phi^{*}({\pr}^{\n}-iA^{\n})\phi
+\l (\phi^{*} \phi-v^{2})^{2}\;\;.     \label{2.1}
\eeq
As Nielsen and Olesen showed \cite{no}, the equations of motion
of (\ref{2.1}) admit stable
vortex string solutions, similar
to those of Abrikosov \cite{abrikosov}, provided that the
higgs mass $m_{H}=2{\sqrt 2\l} v$
is greater than the vector boson mass $m_{VB}=ev$. Recall that
the coherence length is $\xi=m_{H}^{-1}$ and the
penetration depth is $D=m_{VB}^{-1}$.

It is convenient to parametrize
the higgs field $\phi$ by density and phase variables, similar
to those of Tomonaga, Bohm et.
al. \cite{coll}, and Gervais and Sakita \cite{bunji},
\beq
\phi(x)=[v+ h(x)]\; e^{i\int_{-\infty}^{x}\;V_{\m}(y)\;dy^{\m} }\;\;.
\label{2.2}
\eeq
The integral is path dependent, but its exponential is not. The density
field $h$ is restricted by $h\ge -v$. The field $V_{\m}(x)$ is
the superfluid velocity \cite{coll}. Under a gauge transformation
$\delta A_{\m} = {\pr}_{\m}\chi$, $\delta V_{\m}= {\pr}_{\m}\chi$
and $\delta h = 0$. This parametrization is essentially the
same as that of Gervais and Sakita \cite{bunji}.

The field $V_{\m}$ is single-valued, but
$\int_{-\infty}^{x}\;V_{\m}(y)\;dy^{\m}$ is multi-valued. The line integral
of $V_{\m}$ around a closed path is an integer multiple of $2\pi$; the integer
is the (oriented) number of vortex world-sheets the path winds around in four
dimensions. This fact expresses $\Pi_{1} (U(1))=Z$. If the union of all
vortex world-sheets
is the (possibly disjoint) oriented
two-dimensional manifold $\Sigma$, Stokes' theorem
implies
\beq
\pr_{\a} V_{\b} - \pr_{\b} V_{\a}
=2\pi\;\int_{\Sigma}\;d {\sg}^{1} \wedge d{\sg}^{2} \;
\delta^{4}(x-z(\sg))\; \eps_{\a \b \rh \tau}\;
\frac{\pr z^{\rh}}{\pr \sg^{1}}
\frac{\pr z^{\tau}}{\pr \sg^{2}}\;n(\sg)\;, \label{2.3}
\eeq
where $\sg^{1}$ and $\sg^{2}$ are the world-sheet (base space)
coordinates,  $z^{\rh}$ are the space-time (target space)
coordinates and $n(\sg)$ is the vorticity (which is a constant
integer on each connected component of $\Sg$).

Writing the lagrangian (\ref{2.1}) in terms of the
collective fields (\ref{2.2}) and the vortex degrees of freedom
(\ref{2.3}) gives, with $n(\sg)=1$,
\begin{eqnarray}
{\cal L}
&=&\frac{1}{4e^{2}}F_{\m \n} F^{\m \n}+\frac{1}{2}
    {\pr}_{\n}h{\pr}^{\n}h+\l h^{2}(h+2v)^{2}
    +\frac{v^{2}}{2} (A_{\m}-V_{\m})(A^{\m}-V^{\m})   \nonumber \\
&+&\frac{1}{2} h(h+2v)(A_{\m}-V_{\m})(A^{\m}-V^{\m}) \nonumber \\
&+&\frac{i}{2} \eps^{\m \n \a \b} \; A_{\m \n}
    \;[\pr_{\a} V_{\b} - \pr_{\b} V_{\a}
    -\pi\;\int_{\Sigma}\;d {\sg}^{a} \wedge d{\sg}^{b} \;
    \delta^{4}(x-z(\sg))\; \eps_{\a \b \rh \tau}\;
    \frac{\pr z^{\rh}}{\pr \sg^{a}}
    \frac{\pr z^{\tau}}{\pr \sg^{b}}]  \nonumber \\
&+& i\omega(h+v)\;\int_{\Sigma}\;d {\sg}^{1} \wedge d{\sg}^{2}\;
\delta^{4}(x-z(\sg))
\;\;.     \label{2.4}
\end{eqnarray}
The new antisymmetric tensor field $A_{\m \n}$ and scalar
field $\omega$ are Lagrange multipliers. This formalism is not
convenient for all purposes because of the restriction
$h \ge -v$. The last term enforces the condition that $\phi=0$
on the world-sheet. Notice that if the vortex degrees of
freedom, $z(\sg)$ are integrated out in the functional
integral, no new terms depending on the fields $h$ or
$V_{\m}$ are induced in the lagrangian. Thus (\ref{2.4})
is completely equivalent to (\ref{2.1}). In the quantum theory, the
measure on the world-sheet field $z(\sigma)$ must be diffeomorphism
invariant and divided by the volume of the diffeomorphism group.

\section{The Higgs Mass as a UV Cut-off}

The lagrangian (\ref{2.4}) is not easy to work with. The
chief goal of this article is to integrate out the fields in the
target space, leaving behind the action of $z(\sg)$ in the base
space. However this goal is quite difficult with (\ref{2.4}). Fortunately
the
difficulty disappears in the london limit, $\l \rightarrow \infty$. In
the usual way of taking
this limit, the penetration depth, $D$, is fixed, while the coherence
length, $\xi$, tends to zero. Now, if the system is regularized
by an ultraviolet
cut-off $\L$, the coherence length will not vanish as the higgs
mass is increased, but instead tend to
$\L ^{-1}$. Thus, by putting such a cut-off
into the theory, one does not expect the world-sheet action
to be very different from that of the usual abelian
higgs model. This observation
is similar to that of Kirkman and Zachos, who noted that the
energy density away from the core
of a magnetic monopole is determined only by one length scale, namely
the vector boson
mass, if the
higgs mass is very large \cite{kirzac}
Furthermore, it will become clear that there is a
self-energy divergence in the string tension (just as the
electromagnetic
self-energy of an electron is divergent), which
needs to be regularized by a non-zero coherence length.

The lagrangian in the london limit is
\begin{eqnarray}
{\cal L}
&=&\frac{1}{4e^{2}}F_{\m \n} F^{\m \n}+
    \frac{v^{2}}{2} (A_{\m}-V_{\m})(A^{\m}-V^{\m})   \nonumber \\
&+&i \eps^{\m \n \a \b} \; A_{\m \n}
    \pr_{\a} V_{\b}
    -\pi\,i\;A_{\m \n}\int_{\Sigma}\;d {\sg}^{a} \wedge d{\sg}^{b} \;
    \delta^{4}(x-z(\sg))\;  \frac{\pr z^{\m}}{\pr \sg^{a}}
    \frac{\pr z^{\n}}{\pr \sg^{b}}  \nonumber \\
&+& \m_{0}\int_{\Sigma}\;d^{2}{\sg}\;{\sqrt{ det\; g}}
\;\;,     \label{3.1}
\end{eqnarray}
where $g_{a b}$ is the induced metric,
\beq
g_{a b}= e_{a} \cdot e_{b}=\sum_{\m}
\;e^{\m}_{a} \,e^{\m}_{b}\;\;. \label{3.2}
\eeq
and $e_{a}$ is the vector field
\beq
e^{\m}_{a}=\pr_{a} z^{\mu}(\sg) \;\;. \label{3.21}
\eeq
The last term is a Nambu action, which comes about because the
energy density of the higgs field between the vortex core (where $h=-v$)
and the surrounding
vacuum region (where $h=0$) is non-zero. The coefficient of this
term, $\m_{0}$, depends on the coherence
length in a nontrivial way.

\section{Duality}

The next step is to integrate out everything but $z(\sg)$ from
the functional integral with action (\ref{3.1}). There is nothing
fancy about doing this. The result is precisely that obtained
by eliminating the fields from the equations of
motion, since (\ref{3.1}) is at most quadratic in the fields in euclidean
space. As an optional intermediate step, one can integrate
out only $A_{\m}$ and $v_{\m}$ leaving behind only the antisymmetric
tensor field $A_{\m \n}$ as well as $z^{\m}$. The field
strength $F_{\a \m \n}$ is the exterior derivative
of $A_{\m \n}$. The lagrangian of this system can be written
\begin{eqnarray}
{\cal L}
&=& \frac{1}{12v^{2}}\; F_{\a \m \n}F^{\a \m \n}
         +\frac{e^{2}}{2}\; (A_{\m \n}-\pr_{\m} B_{\n}
+\pr_{\n} B_{\n}) (A^{\m \n} -\pr^{\m} B^{\n}
+\pr^{\n} B^{\n}) \nonumber \\
&-& \pi\,i\;A_{\m \n}\int_{\Sigma}\;d {\sg}^{a} \wedge d{\sg}^{b} \;
    \delta^{4}(x-z(\sg))\;  \frac{\pr z^{\m}}{\pr \sg^{a}}
    \frac{\pr z^{\n}}{\pr \sg^{b}}   \nonumber \\
&+& \m_{0}\int_{\Sigma}\;d^{2}{\sg}\;{\sqrt{det\;g}} \;\;. \label{4.1}
\end{eqnarray}
This lagrangian has the gauge invariance $\delta A_{\m \n} =\pr_{\m} \chi_{\n}
-\pr_{\n} \chi_{\n}$, $\delta B_{\m}=\chi_{\m}$. A unitary
gauge exists for which $B_{\m}=0$. Equation (\ref{4.1})
is the dual lagrangian to (\ref{3.1}). It was found
by precisely the same technique used to find the Kramers-Wannier dual
of a lattice field theory or spin system
(see for example \cite{savit}). The duality
of gauge-invariant fields in more than two dimensions
was first understood
by Wegner (who only considered $Z_{2}$ invariant lattice
models. However, Wegner's
method is essentially the same as that used for continuum
abelian systems) \cite{wegner}. The first explicit
discussion in the literature
of the duality of continuum
gauge fields coupled
to continuum scalar fields
appears to be that of Sugamoto \cite{sug}. The
coupling of antisymmetric tensor gauge fields to
strings was first
investigated by Kalb and Ramond \cite{kr}. Kalb and Ramond's model
was rederived and argued to describe
vortices in a relativistic superfluid
by Lund and Regge \cite{regge}. A study of the
radiation of goldstone bosons by global
cosmic strings in the Kalb-Ramond
formulation was made by Vilenkin and Vachaspati \cite{vv}. Nambu
made some suggestions concerning
quark confinement and strings by introducing
a mass in the Kalb-Ramond model \cite{nambu}
and a model of electric
strings in QCD was developed using a non-abelian generalization
of antisymmetric tensor gauge fields on
the lattice \cite{plaq}. Such non-abelian
fields were also discussed by Nepomechie \cite{rafael}. It was
shown that strings are confined by dynamical membranes in Kalb-Ramond
theories with magnetic-monopole instantons, both analytically \cite{atgf}
and numerically \cite{pearson}.

The propagator of the massive antisymmetric tensor gauge field in the
unitary gauge can be written
\beq
S(x-y)^{\m \n ; \a \b}
=\frac{v^{2}}{2(-\pr^{2}+e^{2}v^{2})}\;
(\delta^{\m [\a} \delta^{\b] \n}-\frac{2}{e^{2}v^{2}} \pr^{[\m }
\delta ^{\n] [\a} \pr^{\b]})\;\delta^{4} (x-y)
\label{4.7}
\eeq

\section{The World-Sheet Action}

The easiest way to find the world-sheet action is by taking unitary
gauge $B_{\m}=0$ and integrating out $A_{\m \n}$. A string
tension will be induced by virtue of the exponential fall-off
of the propagator \cite{nambu}, \cite{plaq}. However, the full induced
action has not heretofore been worked out.

Before proceeding further, it is neccesary to
say a bit about the geometry of
two-dimensional manifolds embedded in $R^{4}$. Define the
antisymmetric tensor field on the world-sheet
\beq
t^{\m \n}=\frac{1}{\sqrt{2\,det\;g}} (e^{\m}_{1}\, e^{\n}_{2}
-e^{\m}_{2}\, e^{\n}_{1}) \;\;,   \label{4.2}
\eeq
which will be called the tangent plane field. The tangent plane
field was first discussed by Balachandran
et. al. \cite{bal} and by Polyakov \cite{polyakov}. Notice that
this field satisfies $t_{\m \n}t^{\m \n}=1$. There are two
four-component normal vectors
$n_{k}$, $k=3,4$, satisfying $n_{k}\cdot e_{a}=0$
and $n_{k}\cdot n_{j}=\delta_{kj}$. The
derivative of the tangent vector $e_{a}$ is
given by
\beq
\pr_{a} e_{a}=\G^{c}_{a b}\, e_{c}
+K^{k}_{a b}\, n_{k}\;\;,  \label{4.3}
\eeq
where
$\G^{c}_{a b}$ is the usual affine connection and $K^{k}_{a b}$ is
the second fundamental form or extrinsic curvature tensor.

The tangent plane field satisfies
\beq
\pr_{a} t^{\a \b} = \frac{1}{\sqrt {2\;det\;g}}
(K^{k}_{a 1} n^{\m}_{k} e^{\n}_{2}
-K^{k}_{a 2} n^{\m}_{k} e^{\n}_{1}
-K^{k}_{a 1} n^{\n}_{k} e^{\m}_{2}
+K^{k}_{a 2} n^{\n}_{k} e^{\m}_{2})   \;\;, \label{4.31}
\eeq
and by virtue of $K^{k}_{ab}=K^{k}_{ba}$,
\beq
\pr_{\a} t^{\a \b}
=\sum_{\alpha}
g^{a c}\,e^{\a}_{c}\, \pr_{a} t^{\a \b}  =0    \;\;. \label{4.32}
\eeq

The euclidean space $R^{4}$ can be reparametrized by the new
coordinates $\sg^{1}$, $\sg^{2}$, $\Omega^{3}$ and $\Omega^{4}$
\beq
x^{\m}=x^{\m} (\sg,\Omega) = z^{\m} (\sg)+ \Omega^{k} n^{\m}_{k}(\sg)\;\;,
\label{4.4}
\eeq
though such a parametrization does not assign unique coordinates to each
point of $R^{4}$. The manifold consisting of the disjoint union of a
covering
of $R^{4}$ defined in this way is called the normal bundle of the
manifold $\Sigma$. The normal bundle has base space $\Sigma$ and
each fiber is isomorphic to $R^{2}$. This parametrization has been
used before by Gervais and Sakita for strings with small penetration
depth and coherence length \cite{bunji} and by Mazur and Nair in
their discussion of QCD string actions \cite{nair}.

Points in $R^{4}$ which are close enough to $\Sigma$ (i.e. for which
$(\Omega^{3})^{2}+(\Omega^{4})^{2}$ is sufficiently small) can be
uniquely parametrized by (\ref{4.4}), and the metric in the
new coordinates is
\beq
G_{a b}(\sg, \Omega)=
\sum_{\m}\;
\frac{\pr x^{\m} }{\pr \sg^{a}} \frac{\pr x^{\m} }{\pr \sg^{b}}\;,\;\;
G_{a k}(\sg, \Omega)=
\sum_{\m}\;
\frac{\pr x^{\m} }{\pr \sg^{a}} n^{\m}_{k}\;,\;\;
G_{k j}(\sg, \Omega)=\delta_{k j}   \;\;. \label{4.5}
\eeq
Notice that on the base manifold $\Sigma$, the $2 \times 2$ sub-block
$G_{a b}$ of this metric
reduces to the induced metric, $G_{a b}(\sg, 0)=g_{a b}$. The delta function
of $R^{4}$ in these coordinates is
\beq
\delta^{4}(x^{\m}(\sg,\Omega)  -  x^{\m}(\xi,\Pi))=
\frac{1}{{\sqrt {det\;G}}} \delta^{2}(\sg-\xi)\, \delta^{2}(\Omega-\Pi)
=\frac{1}{{\sqrt {det\;g}}} \delta^{2}(\sg-\xi)\, \delta^{2}(\Omega-\Pi)\;\;.
\label{4.6}
\eeq
which is needed to obtain the propagator (\ref{4.7}) in the new coordinate
system, $\sg^{1}$, $\sg^{2}$, $\Omega^{3}$, $\Omega^{4}$.

One more formula is needed to obtain the world-sheet action in managable
form. For any differential operator $A$ which commutes with $\Omega^{3}$ and
$\Omega^{4}$,
\beq
\int \,d^{2}\Omega\; \delta^{2}(\Omega) \;(-\sum_{k=3}^{4}
\frac{\pr^{2}}{\pr (\Omega^{k})^{2}} +A)^{-1}\;
\delta^{2}(\Omega)
=\frac{1}{4\pi} \int_{0}^{\infty} \,du \;(u+A)^{-1} \;\; . \label{5.2}
\eeq
However, the right-hand-side is ultraviolet divergent. The divergence
can be regularized by introducing an ultraviolet cut-off through
a subtraction procedure. If $M$ is some large number
with the dimensions of inverse centimeters, the regularized
version of (\ref{5.2}) is
\begin{eqnarray}
&REG&\;[\int \,d^{2}\Omega\; \delta^{2}(\Omega) \;(-\sum_{k=3}^{4}
      \frac{\pr^{2}}{\pr (\Omega^{k})^{2}} +A)^{-1}\;
      \delta^{2}(\Omega)]   \nonumber \\
&=  &\frac{1}{4\pi} \int_{0}^{\infty} \,du \;[(u+A)^{-1}
    -(u+M^{2}+A)^{-1}]
=\frac{1}{4\pi}\;\log  \frac{A+M^{2}}{A} \;\; . \label{5.21}
\end{eqnarray}

The world-sheet action obtained from (\ref{4.1}) is
\begin{eqnarray}
I[z]
&=     &\frac{ \pi^{2}}{4}
         \int_{\Sigma}\,d^{2}\sg\;\int\,d^{2}\Omega\;
        \int_{\Sigma}\,d^{2}\xi \,
        \int\,d^{2}\Pi
        \; \sqrt{det\;g(\sg)}\; t^{\m \n}(\sg)\;\delta^{2}(\Omega) \nonumber \\
&\times&S(x(\sg,\Omega)-x(\xi,\Pi))^{\m \n ; \a \b}\;\sqrt{det\;g(\xi)}\;
t^{\a \b}(\xi) \;\delta^{2}(\Pi)\;\;.  \label{5.3}
\end{eqnarray}
This form, equation (\ref{5.3}), of
the world-sheet action is essentially analogous to the
action discussed by Rasetti and Regge, Davis and Shellard
and Lund for superfluid helium \cite{lr}. Using
(\ref{4.7}) together with (\ref{4.6}) for the propagator, as
well as (\ref{4.32}) and (\ref{5.21}), this simplifies to
\begin{eqnarray}
I[z]
&=&\frac{\pi v^{2}}{4}
            \int_{\Sigma} d^{2}\sg \; \sqrt{det\; g} \;\;t^{\m \n}\;
            [\log (1-\frac{\Delta}{\L^{2}})
            - \log (1-\frac{\Delta}{e^{2}v^{2}})]
         \;t_{\m \n}    \nonumber  \\
&+& [\m_{0}+4\pi v^{2} \log (\frac{\L^{2}}{e^{2}v^{2}})]
\int_{\Sigma} d^{2}\sg \;  \sqrt{det\; g}  \;\;, \label{5.4}
\end{eqnarray}
where $\Delta$ is the covariant Laplacian,
\beq
\Delta=\frac{1}{\sqrt{det\;g}}\;
\pr_{a}\,g^{a b} \sqrt{det\; g}\; \pr_{b}\;\;,
\label{5.5}
\eeq
and $\L$ is defined by
$M^{2}=\L^{2}+e^{2}v^{2}$. The action (\ref{5.4}), which is the main
result of this paper, will
henceforth be referred to as the Nielsen-Olesen
action. It is not
a local action, for it is non-polynomial in derivatives.

Physically, one can understand $\L$ as cutting off the high-momentum
modes of the spin-$1$ boson field (formally, this has been viewed
as the dual antisymmetric tensor field, rather than as a vector
field). Thus, it performs the same function as the higgs
mass. Recall that the role of the higgs
field is to ``soften" the high-energy exchange of massive spin-$1$
bosons. The cut-off $\L$ is the inverse coherence length, $\xi^{-1}$, and
its
role, as
far as vortices are concerned, is the same as that of a finite-mass
higgs field.

In the quantum
theory there is another
term which must be added to
(\ref{5.4}); the Liouville action, which has no extrinsic
curvature dependence \cite{poly}.

The validity of the Nielsen-Olesen action (\ref{5.4}) obtained above breaks
down as soon as any eigenvalue of the matrix
$K^{k}$ exceeds the reciprocal
of the penetration depth $D^{-1}=ev$. If this is the case, the normal
bundle is not locally isomorphic to $R^{4}$ a distance $D$ away from the
surface $\Sigma$.

Using (\ref{4.31}), the Nielsen-Olesen action (\ref{5.4}) can be expanded
in powers of the second fundamental form. To leading order
\begin{eqnarray}
I[z]
&\approx& -\frac{\pi v^{2}}{4}
           \; (\frac{1}{e^{2} v^{2}}-\frac{1}{\L^{2}})
          \;\int_{\Sigma} d^{2}\sg \; \sqrt{det\; g} \;\sum_{k}
          \;g^{a b}g^{e d} K^{k}_{a b} K^{k}_{e d} \nonumber \\
&+      & [\m_{0}+4\pi v^{2} \log (\frac{\L^{2}}{e^{2}v^{2}})]
\int_{\Sigma} d^{2}\sg \;  \sqrt{det\; g}  \;\;, \label{5.6}
\end{eqnarray}
hence the string appears to be {\em anti-rigid} at this
order. One might therefore guess that the Nielsen-Olesen
action is unstable. It will be shown, at least for
a special case, that the Nielsen-Olesen action really is stable, so
long as the penetration depth $D$ is greater than the
coherence length $\xi$. However, no
general proof of stability will be given here. Even
for actions with a positive
quadratic term in $K$, some simple
classical solutions are known to be unstable \cite{bz}. In any
case, (\ref{5.6}) is useless, though (\ref{5.4}) is not.

In
the next section it will be shown that a wave instability
occurs if the penetration
depth $D=(ev)^{-1}$ is smaller than the coherence length, $\xi$, which
is equal to the
inverse cut-off $\L^{-1}$. Otherwise, the Nielsen-Olesen action is
stable, at least
around large flat world-sheets. If $D< \L^{-1}$, the
vortex core, with $<\phi>=0$, begins to fluctuate
throughout the volume, destroying the condensate, where
$|\phi|=v$. Thus, the
superconductor type-II can no longer tolerate the
entry of magnetic flux and
becomes type-I. There is a line of phase transitions in the
$D$ vs. $\L^{-1}$ plane at $D=\L^{-1}$, seperating
these two regimes of superconductivity.

The
string tension diverges in the limit of infinite $\L$ (just as the classical
electron self-energy diverges if the electron radius
is set to zero). Therefore, in this limit, the Nielsen-Olesen action
is dominated by the Nambu term.

In references \cite{A}, \cite{A'}, \cite{B}, the ratio
of the coherence length and
the penetration depth is fixed, leaving a single length
scale, $\epsilon$. Perturbation theory is then done
about $\epsilon=0$. The reader can see that at $\L=\infty$
and $e^{2} v^{2}=\infty$ there is no extrinsic
curvature dependence. If one takes $\L=\epsilon^{-1}$ and
$e v=(C \epsilon)^{-1}$ and expands the
Nielsen-Olesen action (\ref{5.4}), the
result is essentially (\ref{5.6}).

\section{Stability of Flat Surfaces }

While the question of stability will not
be completely settled here, it appears likely
that the
Nielsen-Olesen action is stable, unless the
penetration depth is less than the coherence length. In this
section this will
be shown for the special case of a flat surface.

By the implicit function theorem, it is always possible (at least
locally) to impose the gauge $z^{1}=\sg^{1}$, $z^{2}=\sg^{2}$. A
nearly flat surface, can be described by
\beq
z^{1}=\sg^{1}\;,\;\;z^{2}=\sg^{2}\;,\;\;z^{3}=f^{3}(\sg)\;,\;\;
z^{4}=f^{4}(\sg)\;\;.   \label{6.1}
\eeq
Defining the Fourier transform of the fluctuation field, $f$, by
\beq
f^{k}(\sg)=
\int\frac{d^{2}p}{(2\pi)^{2}} \;f^{k}(p) \;e^{-i\;p\cdot \sg}
\;\;, \label{6.2}
\eeq
the
Nielsen-Olesen action (\ref{5.4}) is, to quadratic order
\beq
I[z]=const.+\frac{\pi v^{2}}{8}
\int\frac{d^{2}p}{(2\pi)^{2}}\;\sum_{k}
 \;f^{k}(p)f^{k}(-p) \;p^{2}\,[\log \frac{p^{2}+\L^{2}}{p^{2}+e^{2}v^{2}}
+2\m_{0}]
\;\;, \label{6.3}
\eeq
The frequency of the fluctuations is therefore positive
for all $p$, provided that the coherence length $\xi=\L^{-1}$ is less
than the penetration depth $D=(ev)^{-1}$. Otherwise, there is an
instability
signaling the onset of type-I superconductivity.

\section{The Branched Polymer Phase of the Abelian Higgs Model}

An obvious question is whether the abelian
higgs model has a non-trivial string, with finite tension, after
renormalization. It will be argued here that this is not
the case. The conclusion is perhaps not surprising, for
a theory which is not asymptotically
free in the ultraviolet and which is likely
to be infrared free. The argument hinges on the fact that
the action of a world-sheet tube monotonically decreases
as a function of the radius.

The main limitation of the arguments presented in this
section is that the Liouville action \cite{poly} is not
being considered. However, the Liouville action has no
extrinsic curvature dependence, and is unlikely
to stabilize the string.

Consider a tube of radius $r$. Take $\sg^{1} \; \epsilon \; [0,2\pi r)$ and
$\sg^{2} \; \epsilon \; (-\frac{L}{2}, \frac{L}{2})$ where $L >> r$. The
world-sheet is then described by
\beq
z^{1}(\sg)=\cos \frac{\sg^{1}}{r} \;,\;\;
z^{1}(\sg)=\sin \frac{\sg^{1}}{r} \;,\;\;
z^{3}(\sg)=\sg^{2} \;,\;\;
z^{4}(\sg)=0\;\;.  \label{7.1}
\eeq
The area of the world-sheet is $A=2\pi r L$. The
action of this configuration can be worked out from (\ref{5.4}). The
result is
\beq
I[z]= \m (r) A=
 [\m_{0}+\frac{\pi v^{2}}{4}
\; \log(\frac{\L^{2}+r^{-2} }{e^{2}v^{2}+r^{-2}})]A\;\;.
\label{7.2}
\eeq
Notice that the action of a thin tube (it should
be assumed that $r>D$ for the
Nielsen-Olesen action (\ref{5.4}) to be valid) is
proportional to its
length. Therefore, in
the euclidean quantum field theory, the
usual energy versus entropy arguments for random surfaces \cite{adfo}
imply that branched polymers dominate the world-sheet. This means
that the
quantum abelian higgs model vortex-world-sheet is dominated by
branched-polymer configurations, and the vortex
string tension will not scale in the limit that the
cut-off is removed.

Not all the details of the quantization
of the world-sheet have been taken into account in this
section. In particular, the Liouville
action has been ignored. However, it appears
unlikely that the dominance of branched
polymers will be affected by the
inclusion of the Liouville action. Therefore the
Nielsen-Olesen vortex almost certainly
does not exist
as a quantum soliton of the field theory.

\section{Discussion}

In this paper, the full world-sheet action of a Nielsen-Olesen
vortex in an abelian higgs model was obtained, where the effect
of a finite higgs mass was approximated by the introduction of
an ultraviolet cut-off. In particular, the extrinsic
curvature dependence was found. The action
was argued to be stable, provided that the
penetration depth is greater than the coherence
length. It is almost certainly the
case that the vortex
world-sheets of the quantum field theory are dominated by branched
polymer configurations, implying that no scaling limit exists in
which the renormalized
string tension is finite.

\section*{Acknowledgements}

I am grateful to Mark Bowick, Dimitra
Karabali, V.~P. Nair, Bunji Sakita and especially
Cosmas Zachos for informative discussions. I thank
M.~I. Polikarpov for informing me of reference
\cite{polikarpov}, in
which some of the techniques discussed
in this paper were used
for the lattice abelian higgs model. Arne
Larsen was very helpful at all
stages of this work. I would finally
like to thank Poul Olesen, who checked the result
(\ref{5.4}) using zeta-function regularization, and
pointed out the relevance of reference \cite{arodz}.

\section*{Note Added}

The world-sheet action
of a non-relativis\-tic vortex can be
obtained using similar methods. After this paper appeared
as a preprint, the author discovered that a number of people
had recently studied such Abrikosov
vortex strings using dual Kalb-Ramond
fields \cite{zee}, \cite{thouless}. In particular, the
action analogous to (8) (originally found in \cite{lr})
has been obtained \cite{thouless} in this way. The
authors of reference \cite{thouless} intentionally dropped
certain terms containing time derivatives, which were not
needed for their purposes. Including these terms, the author
has reduced the action for a nonrelativistic vortex
to an
action in which the dependence on the
curvature of the vortex is made explicit, using
the normal-bundle parametrization discussed here. It
can then be shown that for this case as well, flat
surfaces are stable provided the coherence length
is less than the penetration depth.

Recently, Sato and Yahikozawa calculated the extrinsic
curvature dependence of the relativistic Nielsen-Olesen world-sheet
to second order in $K$, obtaining a result similar to (22) \cite{sy}. I
would like to thank them for discussions concerning their work.

\vfill

\end{document}